# Improper legitimization of hijacked journals through citations[1]

Anna Abalkina, Freie Universität Berlin, Garystraße 55, Berlin, 14195, Germany
anna.abalkina@fu-berlin.de

Guillaume Cabanac, University of Toulouse, Computer Science Department, IRIT UMR 5505 CNRS, 31062 Toulouse, France, guillaume.cabanac@univ-tlse3.fr

Cyril Labbé, Univ. Grenoble Alpes, CNRS, Grenoble INP, LIG, 38000 Grenoble, France, cyril.labbe@univ-grenoble-alpes.fr

Alexander Magazinov, Yandex, 82 Sadovnicheskaya str., Moscow 115035, Russia, magazinov-al@yandex.ru

## Objective

Hijacked journals are cyber-criminal publishers that clone the title, ISSN and other metadata of a legitimate journal. Hijacked journals collect fees for the publication without performing peer review. There are more than 200 documented cases of hijacked journals [1]. The published results are often based on low-quality or dishonest research practices. The goal is to study the impact of articles published in hijacked journals on authentic scientific journals through citations. Moussa [2] limited the analysis of citations of articles in hijacked journals to the field of marketing. Do references to articles originating from hijacked journals infiltrate scientific communication? This study raises awareness of the risk of legitimizing unreliable articles and fraudulent journals.

## Design

This study covered all records indexed by Dimensions, a bibliometric database containing more than 100 million publications. A 'Citejacked detector' was designed as a part of the *Problematic Paper Screener* (https://www.irit.fr/~Guillaume.Cabanac/problematic-paper-screener). It tabulated questionable articles citing hijacked journals, i.e.'citejacked

---

articles'. The [detector](#) screened 12 journals that were documented to be hijacked, with the hijacked version erroneously indexed in international bibliographic databases. A full-text search was performed between November 2021 and January 2022 in the Dimensions database using the name of 1 of the 12 hijacked journals and included articles published and indexed in Dimensions between 01.Jan.2021-31.Jan.2022. To exclude items published in preprints or predatory journals, the search was limited to the articles published in journals listed in the 'Norwegian Register for Scientific Journals, Series and Publishers'. The manual check of each retrieved item to retain true positives (citations to hijacked journals) and discard false positives (citations to authentic journals or other mentions of journals) was performed.

## Results

This method retrieved 1,421 articles featuring the name of a hijacked journal. The manual analysis of the bibliography in these articles revealed that 828 (58,3%) of them cite unreliable articles from hijacked journals. Citejacked articles were published by 67 publishers. Figure 1 showed the distribution of citejacked articles per publisher: flagship publishers were not immune to incorporating references to hijacked journals into their citation index. During 01.Jan.2021-31.Jan.2022, an average of 2 citejacked articles has been published daily in established journals.

## Conclusions

The results showed that citations of unreliable articles from hijacked journals in reputable journals legitimized such pseudo-articles. The presence of citejacked articles showed that curation of references must be taken more seriously and could speak of a flaw in a peer review process. These flawed references to illegitimate journals may serve as predictors of problematic articles due to plagiarism in articles originating from hijacked journals, citation cartels, and paper mills. Given the limited number of titles included in this study, 12 among the 200+ ever documented [1], the phenomenon might be wider and is not yet systematically studied.

Figure 1: Number and share of citejacked articles per publisher (top 10 only) between January 2021 and January 2022

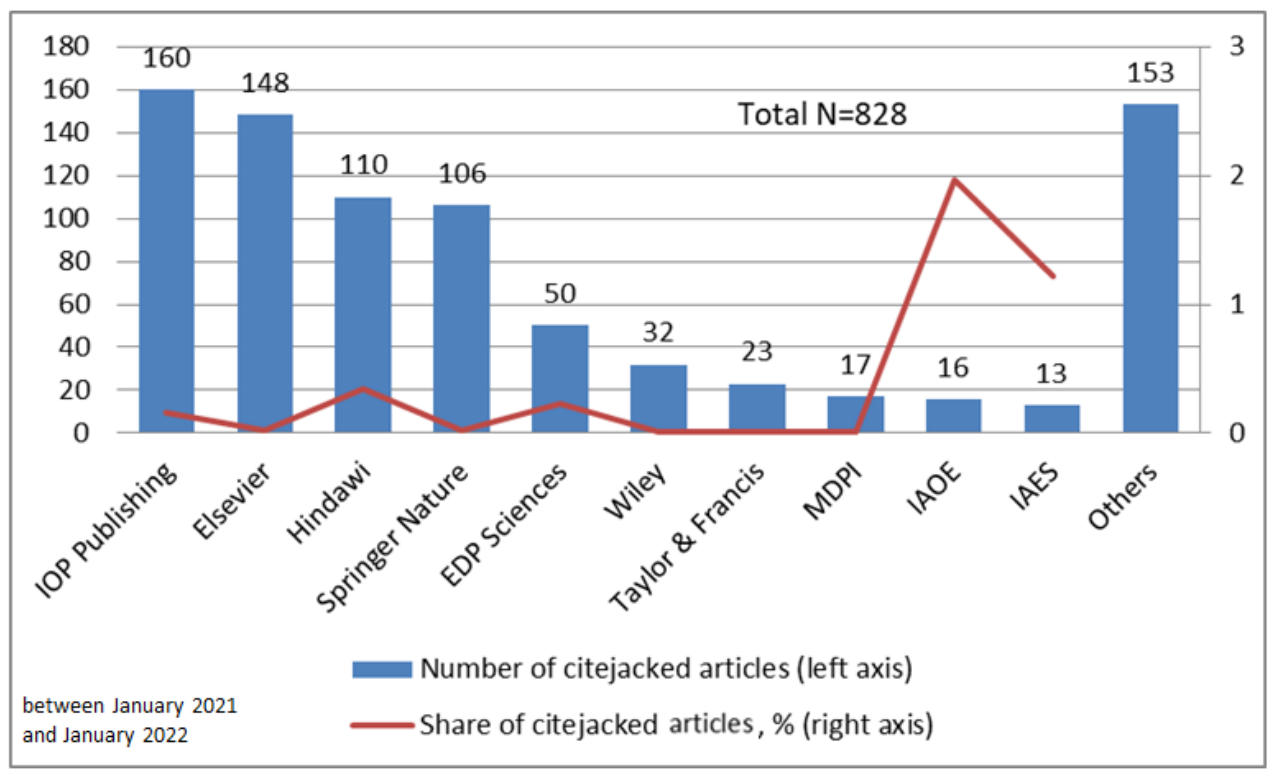

Conflict of Interest Disclosures: none to report

Funding/Support:
Cyril Labbé and Guillaume Cabanac: ERC-Synergy NanoBubbles (ERC-2020-SyG-951393)